\documentclass[tighten, trackchanges]{aastex631}
\accepted{to the Astronomical Journal}
\usepackage{graphicx}
\usepackage{natbib}
\usepackage{latexsym}
\usepackage{amssymb}
\usepackage{longtable}
\usepackage{amsmath}
\usepackage{url}

\def\msun{\ifmmode {\rm\,M_\odot}\else ${\rm\,M_\odot}$\fi}
\def\Msun{\ifmmode {\rm\,\it{M_\odot}}\else ${\rm\,M_\odot}$\fi}
\def\lsun{\ifmmode {\rm\,L_\odot}\else ${\rm\,L_\odot}$\fi}
\def\Lsun{\ifmmode {\rm\,\it{L_\odot}}\else ${\rm\,L_\odot}$\fi}
\def\rsun{\ifmmode {\rm\,R_\odot}\else ${\rm\,R_\odot}$\fi}
\def\Rsun{\ifmmode {\rm\,\it{R_\odot}}\else ${\rm\,R_\odot}$\fi}
\def\Tsun{\ifmmode {\rm\,T_\odot}\else ${\rm\,T_\odot}$\fi}
\def\arcsec{\ifmmode {^{\prime\prime}}\else $^{\prime\prime}$\fi}
\def\asec{\ifmmode {^{\prime\prime}}\else $^{\prime\prime}$\fi}
\def\arcmin{\ifmmode {^{\prime}}\else $^{\prime}$\fi}
\def\amin{\ifmmode {^{\prime}}\else $^{\prime}$\fi}
\def\simlt{\mathrel{\spose{\lower 3pt\hbox{$\mathchar"218$}}
     \raise 2.0pt\hbox{$\mathchar"13C$}}}
\def\simgt{\mathrel{\spose{\lower 3pt\hbox{$\mathchar"218$}}
\     \raise 2.0pt\hbox{$\mathchar"13E$}}}

\def\Snospace~{\S{}}

\newcommand{\lasp}{Laboratory for Atmospheric and Space Physics, University of Colorado, 600 UCB, Boulder, CO 80309, USA}
\newcommand{\casa}{Center for Astrophysics and Space Astronomy, University of Colorado, 593 UCB, Boulder, CO 80309}
\newcommand{\cuboulder}{Department of Astrophysical and Planetary Sciences, University of Colorado, Boulder, CO 80309, USA}
\newcommand{\iowa}{The University of Iowa, Department of Physics \& Astronomy, Van Allen Hall, Iowa City, IA 52242, USA}

\newcommand{\vanderbilt}{Department of Physics and Astronomy, Vanderbilt University, Nashville, TN 37235, USA}

\usepackage{xcolor}
\usepackage[normalem]{ulem}
\newcommand\redsout{\bgroup\markoverwith{\textcolor{red}{\rule[0.5ex]{2pt}{0.4pt}}}\ULon}

\begin{document}

\author[0000-0002-7119-2543]{Girish M. Duvvuri}
\affiliation{\cuboulder}
\affiliation{\casa}
\affiliation{\lasp}
\affiliation{\vanderbilt}

\author[0000-0001-9207-0564]{P. Wilson Cauley}
\affiliation{\lasp}

\author[0000-0003-4628-8524]{Fernando Cruz Aguirre}
\affiliation{\cuboulder}
\affiliation{\lasp}
\affiliation{\iowa}

\author{Roy~Kilgard}
\affiliation{Astronomy Department, Wesleyan University, Middletown, CT 06459, USA}

\author[0000-0002-1002-3674]{Kevin France}
\affiliation{\cuboulder}
\affiliation{\lasp}
\affiliation{\casa}

\author[0000-0002-3321-4924]{Zachory K. Berta-Thompson}
\affiliation{\cuboulder}
\affiliation{\casa}

\author[0000-0002-4489-0135]{J. Sebastian Pineda}
\affiliation{\lasp}

\correspondingauthor{Girish M. Duvvuri}
\email{girish.duvvuri@colorado.edu}

\title{The High-Energy Spectrum of the Young Planet Host V1298 Tau}

\shorttitle{The High-Energy Spectrum of the Young Planet Host V1298 Tau}

\begin{abstract}
V1298 Tau is a young pre-main sequence star hosting four known exoplanets that are prime targets for transmission spectroscopy with current-generation instruments. This work pieces together observations from the \emph{NICER} X-ray telescope, the Space Telescope Imaging Spectrograph and Cosmic Origins Spectrograph instruments aboard \emph{Hubble Space Telescope}, and empirically informed models to create a panchromatic spectral energy distribution for V1298 Tau spanning 1 -- $10^5$ \AA. We describe the methods and assumptions used to assemble the panchromatic spectrum and show that despite this star's brightness, its high-energy spectrum is near the limit of present X-ray and ultraviolet observatories' abilities to characterize. We conclude by using the V1298 Tau spectrum as a benchmark for the activity saturation stage of high-energy radiation from solar-mass stars to compare the lifetime cumulative high-energy irradiation of the V1298 Tau planets to other planets orbiting similarly massive stars.
 
\end{abstract}

\keywords{}

\section{INTRODUCTION}
\label{sec:intro}

V1298 Tau is a pre-main sequence star that hosts 4 known transiting exoplanets \citep{David:2019a, David:2019b}. The star is bright ($d=108.5$ pc, $m_{\textrm{Gaia}} = 10.1$, \citealp{GaiaCollaboration:2018}) and similar to the young Sun ($M_{\star} = 1.101 M_{\odot}$, $R_{\star} = 1.345 R_{\odot}$, spectral type between K0 -- K1.5, $23 \pm 4$ Myr old, \citealp{David:2019b}), making the V1298 Tau planets prime targets for transmission spectroscopy. Both the star and its planets will change significantly over the lifetime of the system: the star will spin down, contract, and emit less high-energy radiation while the planets will contract as they both cool and lose mass from their H/He envelopes. The majority of planetary atmospheric escape is expected to take place within the first Gyr of the system's lifetime \citep{King:2021} and studying the physics of atmospheric evolution is necessary to understand exoplanet demographics and habitability. A major open question in this area is whether formation conditions or evolutionary processes like photoevaporative mass loss \citep{Watson:1981} and core-powered heating \citep{Ginzburg:2018} are primarily responsible for the ``radius valley": an apparent sparsity of exoplanets with radii near 1.8 $R_{\oplus}$ \citep{Fulton:2017}. Statistical experiments have been proposed to compare predictions from both atmospheric loss mechanisms to the observed exoplanet population, but these approaches rely on input assumptions of the initial high-energy fluxes of young stars and their subsequent evolution \citep{Rogers:2021}. Determining the high-energy irradiation and atmospheric escape of young exoplanets like those orbiting V1298 Tau is necessary to assess the accuracy and precision of those input assumptions, following through to how we understand early planet atmospheres in our Solar System and beyond.

\begin{figure}
    \centering
    \includegraphics[width=\textwidth]{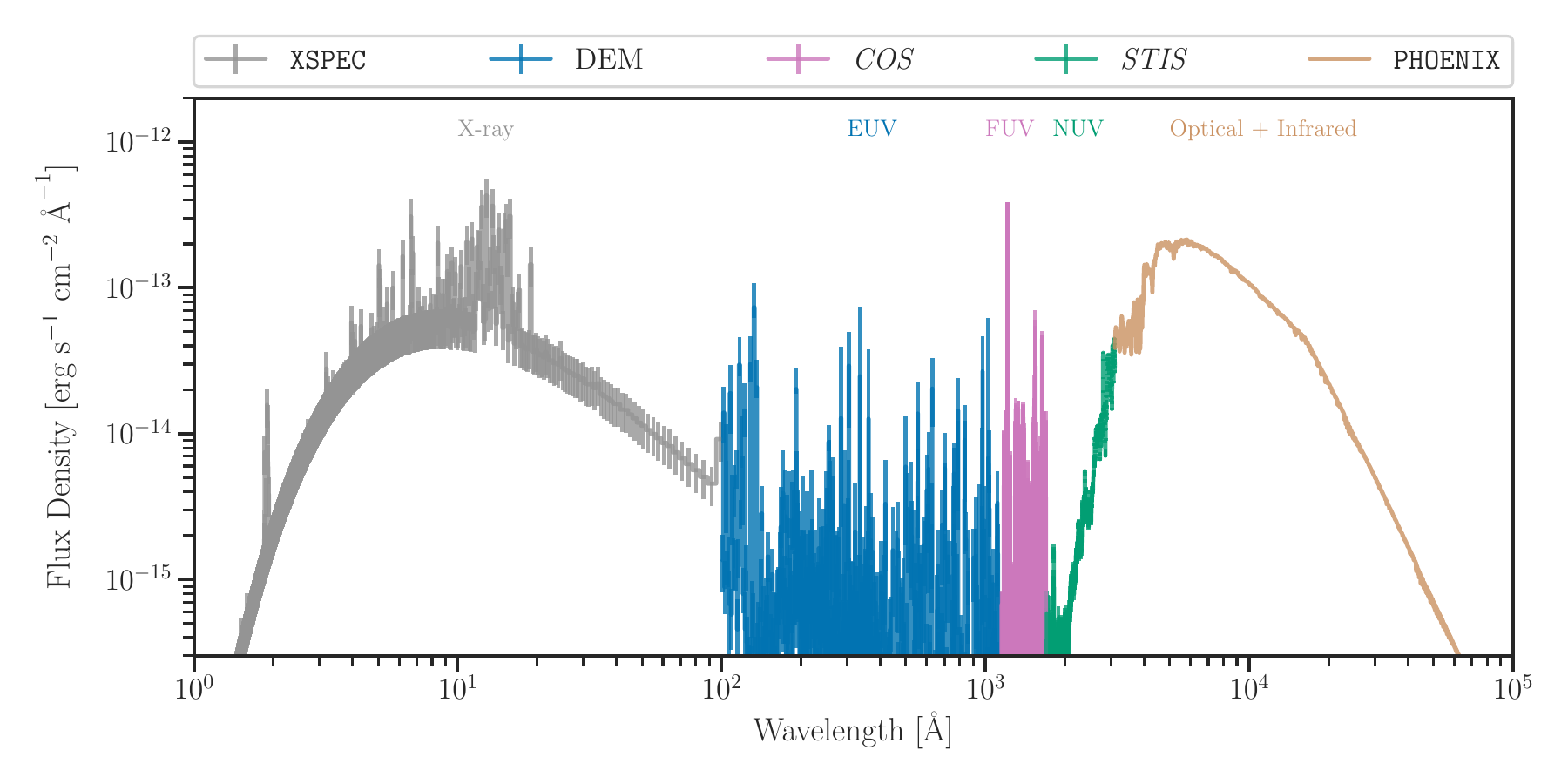}
    \caption{The composite spectrum is plotted with each component covering a specific wavelength interval at its original wavelength resolution, using data where available and supplemented by empirically constrained models. The components and their respective wavelength intervals are: XSPEC model (gray), 1 -- 100 \AA;  Differential Emission Measure model (light blue), 100 -- 1150 \AA; \textit{HST} COS data (pink), 1150 -- 1700 \AA, with two sub-intervals 1214.63 -- 1216.78 \AA\ (Lyman-$\alpha$) and 1519.42 -- 1530.78 \AA\ replaced with scaled excerpts from the MUSCLES SED for $\epsilon$ Eridani; \textit{HST} STIS data (green), 1700 -- 3100 \AA; PHOENIX model (light brown), 3100 -- 10$^5$ \AA.}
    \label{fig:full_spec}
\end{figure}

This work describes the creation of a panchromatic spectral energy distribution (SED) for V1298 Tau (wavelengths from $1$ -- 10$^5$ \r{A}) made available as a data product for the community to use when modeling this planetary system and interpreting observations of atmospheric escape. The panchromatic SED is presented in Figure \ref{fig:full_spec}. \autoref{sec:obs} lists the X-ray and ultraviolet observations contributing to the spectrum, \autoref{sec:analysis} describes our analysis of the star's far ultraviolet (FUV, 1140 -- 1710 \r{A}) emission lines and coronal properties, and \autoref{sec:euv} explains the method used to predict the unobserved extreme ultraviolet (EUV, 100 -- 912 \AA) flux and compares this work's inferred EUV flux to similar work by \citet{Poppenhaeger:2021} and \citet{Maggio:2023}. \autoref{sec:conclusion} concludes by using the V1298 Tau spectrum to characterize the lifetime high-energy irradiation of planets orbiting solar-mass stars.

\section{Observations}
\label{sec:obs}

From 2020 through early 2022 we obtained observations of V1298 Tau's high-energy spectrum using the Space Telescope Imaging Spectrograph (STIS, \citealp{Woodgate:1998}) and Cosmic Origins Spectrograph (COS, \citealp{Green:2012}) instruments on the \textit{Hubble Space Telescope (HST)} and NASA's \textit{Neutron Star Interior Composition ExploreR (NICER)} mission aboard the International Space Station \citep{Gendreau:2016}. The ultraviolet observations cover the wavelength range 1140 \AA\ -- 3150 \AA\ and the X-ray observations span the energy range 0.1 -- 10 keV ($\approx 5$ \AA\ - 55 \AA). We detail the individual instrument settings and observations in the two subsections below, including a summary in \autoref{tab:obs}. 

\begin{deluxetable*}{lccccccc}
\tablecaption{Summary of \textit{NICER} and \textit{HST} observations\label{tab:obs}}
\tablehead{\colhead{}&\colhead{}&\colhead{Date}&\colhead{Starting time}&\colhead{Exposure time}&\colhead{$\lambda_\text{start}$}&\colhead{$\lambda_\text{end}$}&\colhead{$\Delta \lambda^\dagger$}\\
\colhead{Telescope}&\colhead{Instrument setting}&\colhead{(UT)}&\colhead{(UT)}&\colhead{(seconds)}&\colhead{(\AA)}&\colhead{(\AA)}&\colhead{(\AA)}}
\colnumbers
\tabletypesize{\scriptsize}
\startdata
 \textit{NICER} & \nodata & 2020-09-13 & 11:06 & 1880 & 5 & 55 & 0.9 \\
 \textit{NICER} & \nodata & 2020-10-18 & 23:37 & 2134 & 5 & 55 & 0.9 \\
 \textit{HST} & STIS G230L & 2020-11-07 & 06:44 & 21924 & 1600 & 3150 & 3.0 \\
 \textit{HST} & COS G160M & 2020-10-17 & 14:54 & 1998 & 1350 & 1710 & 0.09\\
 \textit{HST} & COS G130M & 2021-12-23 & 10:44 & 9892 & 1140 & 1420 & 0.09\\ 
  &  & 2022-01-17 & 03:18 & 12030 & 1140 & 1420 & 0.09 \\
\enddata
\tablenotetext{^\dagger}{Resolutions vary across the free spectral range. We report the approximate value at the central wavelength of the recorded spectrum.}
\end{deluxetable*}

\subsection{Hubble Space Telescope}
\label{sec:hst}

The \textit{HST} observations (GO 16163, PI~--~P. Cauley) were designed to span the FUV and near ultraviolet (NUV, 1710 -- 3150 \r{A}) spectral ranges with minimal gaps in coverage. To accomplish this we utilized two COS settings and a single STIS setting. The COS observations were obtained with the G130M and G160M gratings and cover the FUV wavelengths and the STIS observations were performed with the G230L grating to cover the NUV spectral range. We note that the COS G130M observations were executed during transits of V1298 Tau c with the goal of measuring mass loss from the planet's atmosphere. The transit observations will be detailed in an upcoming paper. Here, we combine the first two out of four G130M visits into a high-quality FUV spectrum to be included in the final SED data product. The transit depth of V1298 Tau c is $< 0.2\%$ \citep{David:2019b} and the presence of transits during the FUV observations has negligible impact on the total line flux measurements from the co-added spectrum.

\subsection{NICER}
\label{sec:nicer}

\textit{NICER} is a soft X-ray telescope whose primary purpose is to investigate the equation of state of the interiors of neutron stars. \textit{NICER} was designed to have high photon arrival time accuracy and is able to record events with a precision of $< 300$ nanoseconds, but its excellent soft X-ray sensitivity also makes it useful for observing the high-energy emission from stellar coronae. \textit{NICER} only has a single configuration so we do not specify the instrument Grating/Setting in \autoref{tab:obs}. We obtained $\approx 4$ ks of exposure time through \textit{NICER}'s Guest Observer Program Cycle 2 (proposal number 3041, PI ~--~ Cauley) on two separate dates: 1880 seconds on 2020-09-13 and 2134 seconds on 2020-10-18.

\section{Analysis}
\label{sec:analysis}

We analyzed the X-ray and FUV data to provide constraints for estimating the EUV spectrum and the intrinsic stellar Lyman-$\alpha$ profile. To complete the panchromatic spectrum beyond the \textit{HST} STIS G230L observations we follow the MUSCLES approach and use a \texttt{PHOENIX} model with $T_{\rm eff} = 5000$ K, $\log g = 4.0$, [Fe/H] $=0.0$ \citep{Husser:2013}, resampled to a wavelength resolution of $~1.5$ \AA\ and rotationally broadened to 23 km s$^{-1}$. After scaling the PHOENIX model to match the STIS data at 3100 \AA, the model and data showed good agreement between 2800 and 3100 \AA, suggesting that this model is a good approximation for this star's spectrum at longer wavelengths. The scaled \texttt{PHOENIX} spectrum component covers 3100 -- $10^5$ \AA. While there is an optical and infrared spectrum of V1298 Tau \citep{Feinstein:2021} from 4000 -- $10^4$ \AA, there is no overlap with the STIS data and aligning the flux calibration of this chunk of the spectrum between portions of the PHOENIX model was beyond the scope of this work.

\subsection{X-ray Analysis}
\label{sec:analysis_xray}
\begin{figure}
    \centering
    \includegraphics[width=\textwidth]{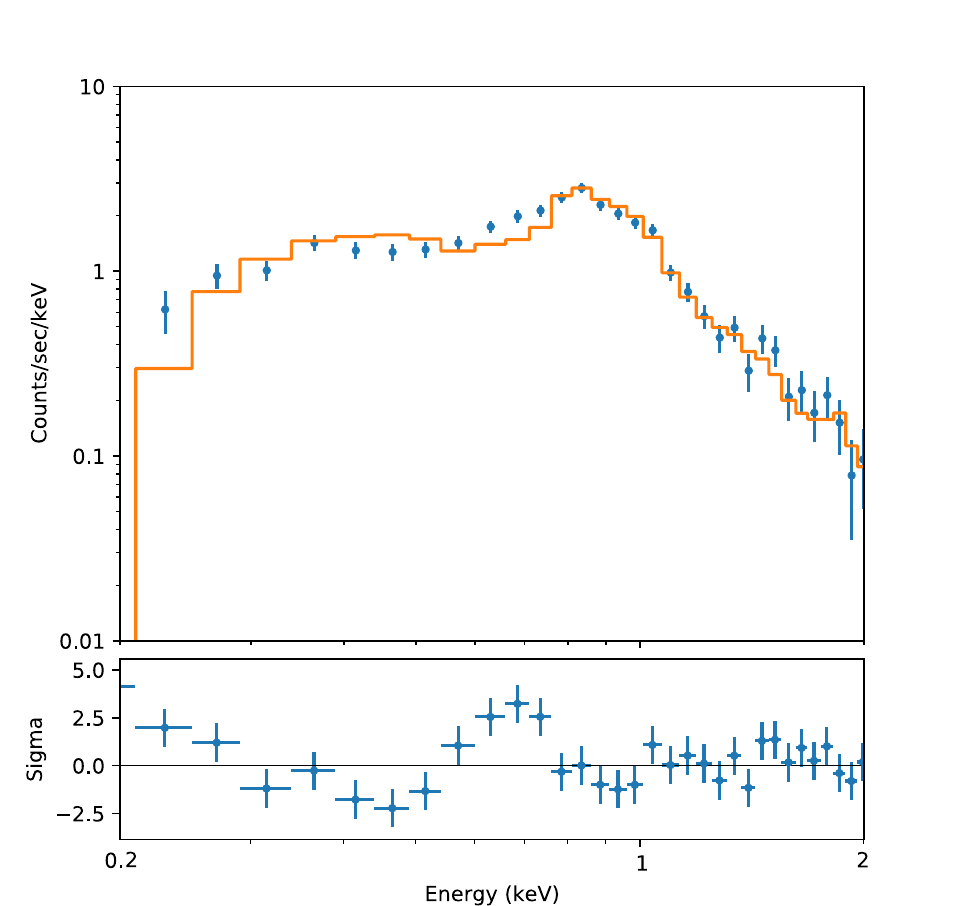}
    \caption{The \textit{NICER} spectrum from the October 2020 observation is plotted as blue circles with errorbars while the best-fit XSPEC model is plotted in solid orange. The model fits the continuum and strong emission lines at intermediate energies well but is less consistent with the emission from low energies.}
    \label{fig:xray}
\end{figure}

We processed both NICER observations using NICERDAS 9/HEASoft 6.30 \citep{HEASARC:2014} to generate cleaned event lists, extract spectra, and generate observation-specific response functions. We estimated the background levels using the nibackgen350 tool of \citet{Remillard:2022} and modeled the spectra in XSPEC \citep{Arnaud:1996} with photoelectric absorption and a Raymond-Smith optically thin thermal plasma model \citep{Raymond:1977}. The spectral fit parameters and fluxes were nearly identical in both NICER observations (see \autoref{tab:xray}), with $n($\ion{H}{1}$) = 2.42 \pm 0.54 \times 10^{20}$ cm$^{-2}$, a plasma temperature of $k_{\textrm{B}}T = 0.79\pm 0.015$ keV, sub-Solar metallicity abundance ($\approx 0.1$), and an observed flux in the 0.1-10 keV band of $1.8 \times 10^{-12}$ erg cm$^{-2}$ s$^{-1}$. A plot of the observed X-ray spectrum and model for the second observation is shown in Figure \ref{fig:xray}. For the final data product we use the XSPEC model for wavelengths below $100 \, \textrm{\AA}$ and adopt a conservative flat uncertainty of 30\% across this component of the SED.

\begin{deluxetable*}{lcccc}
\tablecaption{Spectral fits to \textit{NICER} observations\label{tab:xray}}
\tablehead{\colhead{NICER ID}&\colhead{Temperature}&\colhead{Abundance}&\colhead{$\chi^2/D.O.F.$}&\colhead{Observed Flux (0.1-10 keV)}\\
\colhead{}&\colhead{(keV)}&\colhead{}&\colhead{}&\colhead{(erg cm$^{-2}$ s$^{-1}$)}}
\tabletypesize{\scriptsize}
\startdata
3541010201 & $0.79\pm 0.015$ & $0.11\pm 0.013$ & 293.7/96 & $1.84\times 10^{-12}$ \\
3541010301 & $0.79\pm 0.014$ & $0.14\pm 0.017$ & 109.54/82 & $1.81\times 10^{-12}$ \\
\enddata
\end{deluxetable*}

\subsection{Far-UV Emission Line Measurements of V1298 Tau}
\label{sec:analysis_fuv}
V1298 Tau was observed with the medium-resolution far-UV modes of COS (G130M and G160M; \citealt{Green:2012}) as part of GO 16163 (PI~--~P. Cauley). These observations  (program ID GO 16163, visits 2, 3, and 4) were acquired between 17 October 2020 and 17 January 2022. G130M observations were acquired in the CENWAVE 1291, FP-POS 4 setting, and G160M observations were acquired in the CENWAVE 1533 setting using all four FP-POS tilts.  Together, these observations create a nearly continuous FUV spectrum from 1140~--~1710~\AA, with an 11~\AA\ gap around 1525~\AA\ where the COS detector segments are physically separated, and mitigate the effects of fixed pattern noise.  The one-dimensional spectra produced by the COS calibration pipeline, CALCOS, were aligned and coadded using the custom software procedure described by~\citet{France:2012}.  The final FUV spectrum has a point-source resolution of $\Delta$ $v$~$\approx$~20 km s$^{-1}$ with 6~--~7 pixels per resolution element. A three-pixel boxcar smoothing was applied prior to fitting the emission lines. The total far-UV exposure times were 21,924s in G130M and 1,998s in G160M.  

\begin{figure}[htbp]

   \centering
    \includegraphics[clip,trim=0mm 0mm 5mm 5mm,angle=0, width=\textwidth]{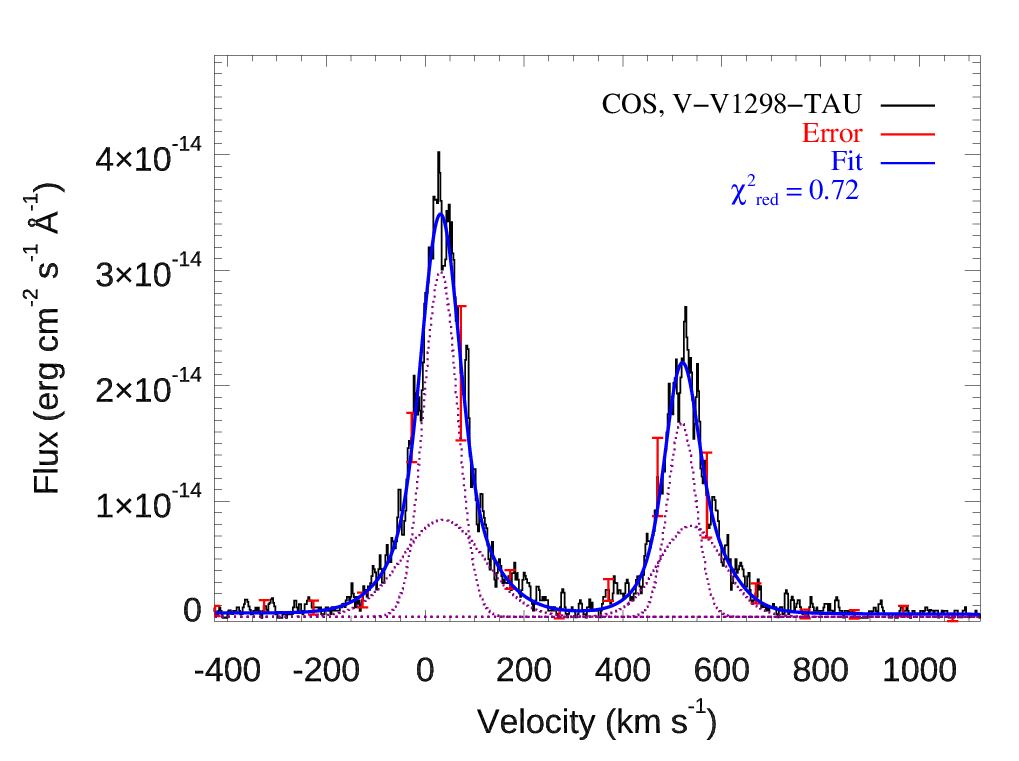}
   \figcaption{The \ion{C}{4} doublet from V1298 Tau.  COS/G160M spectra are shown as the black histogram, with representative error bars in red.  A two-component Gaussian fit is shown overplotted; individual components are in the dashed magenta lines and the overall fit is in solid blue. \label{fig:fuv_line_fit}}
\end{figure}

The chromospheric, transition region, and coronal emission lines in the COS spectra were fitted with an interactive multi-Gaussian line-fitting code optimized for COS emission line spectra. This code assumes a Gaussian line-shape convolved with the wavelength dependent line-spread function, then uses the MPFIT routine to minimize $\chi^{2}$ between the fit and data~\citep{Markwardt:2009, France:2012}. A second order polynomial background, the Gaussian amplitudes, and the Gaussian full-widths-at-half-maximum (FWHM) for each component are free parameters. The parameters of the underlying Gaussian emission lines are returned to the user, and the total line fluxes (\autoref{tab:lines}) are used as inputs to the DEM calculations described in \autoref{sec:dem}. \autoref{fig:fuv_line_fit} presents the spectrum and line fit for the \ion{C}{4} emission line as an example of the data and line-fitting procedure.

\subsection{Lyman-alpha Recovery}
\label{sec:lyman}
Stellar Lyman-$\alpha$ emission is obscured by \ion{H}{1} in the interstellar medium (ISM) which attenuates the line core. Observing Lyman-$\alpha$ with \textit{HST}, whose orbit lies within the Earth's exosphere, is further complicated by geocoronal Lyman-$\alpha$ emission, otherwise referred to as airglow. For COS data, the airglow signal cannot be separated from the stellar signal during the standard background subtraction routine. \citet{CruzAguirre2023} (hereafter referred to as CA23) developed a tool which subtracts airglow emission from COS data to recover the underlying stellar Lyman-$\alpha$ emission by simultaneously fitting the intrinsic stellar emission, ISM absorption, and the contaminating airglow. While the tool was designed for main sequence F-, G-, K-, and M-type dwarf stars in the stellar neighborhood ($\lesssim$ 80 pc), we attempted to use the tool to recover the faint Lyman-$\alpha$ emission of V1298 Tau. Due to the distance to V1298 Tau being larger than what the tool was optimized for, we increased the maximum \ion{H}{1} column density to 10$^{20}$ cm$^{-2}$, based on measured column densities at similar distances being $\sim 10^{19.6}$ cm$^{-2}$ \citep{Wood2005}. The spectral location of the airglow profile changes over time due to the motion of the spacecraft and the time elapsed between COS observations was large enough to require separate airglow subtractions for each individual observation. 

\begin{figure}
    \centering
    \includegraphics[width=\textwidth]{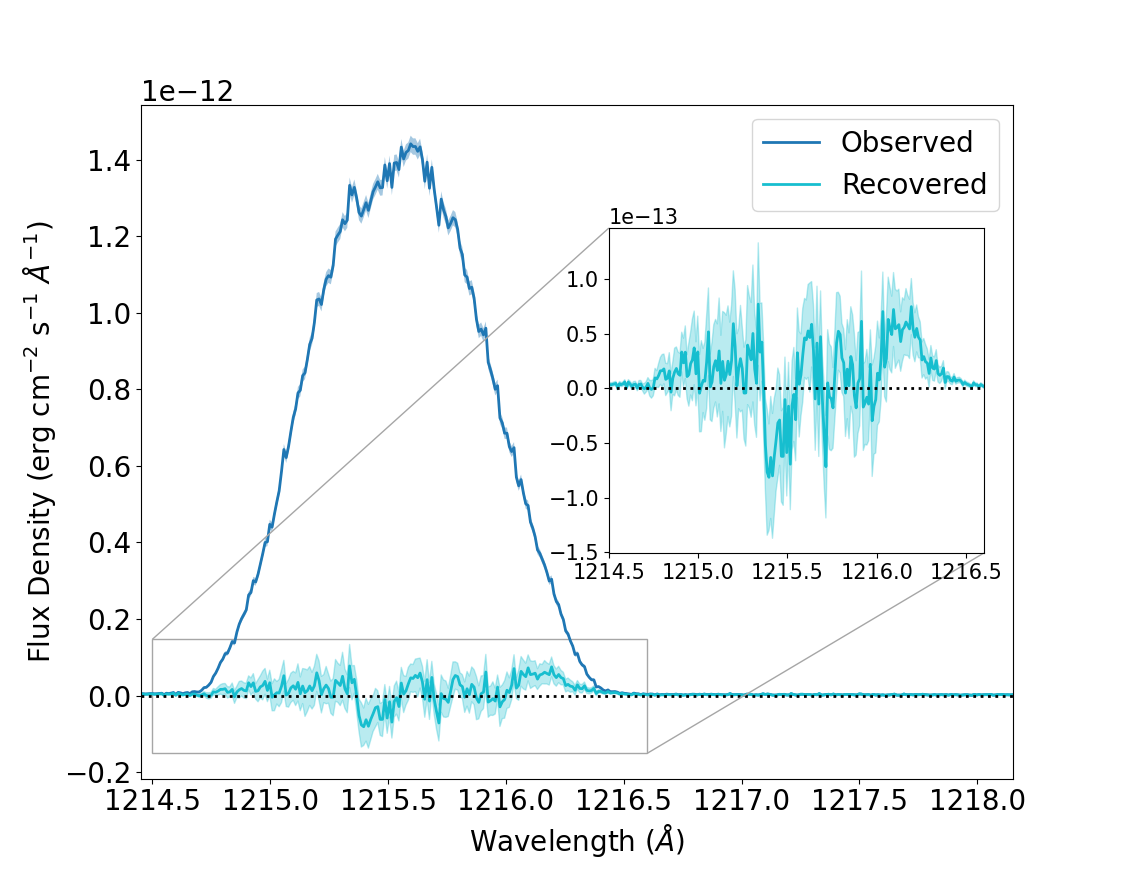}
    \caption{Lyman-$\alpha$ airglow subtraction of V1298 Tau. The spectrum as observed by COS is shown in dark blue. The CA23 tool is used to subtract the airglow, resulting in the recovered (ISM attenuated) spectrum in light blue. The recovered signal of V1298 Tau is faint, and a reliable reconstruction of the stellar emission was not possible.}
    \label{fig:lya}
\end{figure}

The contaminating airglow dominates the observed spectrum, as shown in Figure \ref{fig:lya}, leaving behind little flux to inform the reconstruction of the intrinsic stellar emission line profile. The retrieval is further complicated by the effects of gain sag on the COS detector in the vicinity of geocoronal Lyman-$\alpha$, which reduces the throughput of the stellar signal and was the primary cause for failed Lyman-$\alpha$ recoveries in CA23. Only two of the three recovered profiles were consistent in their shape, and were co-added together to try to improve the quality of the fit, but the results were poorly constrained and unstable even after multiple simplifications to the model constraining the intrinsic line profile. Therefore we elected to estimate the Lyman-$\alpha$ flux of V1298 Tau using empirically calibrated scaling relations.

There are multiple correlation methods to predict the integrated Lyman-$\alpha$ flux using other more accessible quantities, divided into either measured fluxes from emission lines or stellar parameters. These correlation methods are calibrated using samples of nearby stars where Lyman-$\alpha$ reconstructions are more viable, but these are typically main-sequence stars. Table \ref{tab:lya_corr} lists the Lyman-$\alpha$ flux predicted by a number of relations available in the literature, each using different activity tracers or proxies. All relations from CA23 and \citet{Wood2005} take the form of a power-law, while the \citet{Pineda:2021a} prediction uses the saturation value of the Lyman-$\alpha$ $\frac{F_{\textrm{Ly}\alpha}}{L_{\textrm{bol}}}$ broken power-law relation because V1298 Tau is a fast enough rotator to be in the saturated regime. We adopt the integrated flux predicted by the \citet{Wood2005} \ion{Mg}{2} relation because the other line-based relations are from transition region lines, formed over a narrower spatial and temperature range than Lyman-$\alpha$.

We chose to scale the Lyman-$\alpha$ reconstruction of $\epsilon$ Eridani from the MUSCLES data products \citep{France2016, Youngblood:2016} because it is the youngest K star with a published high-quality Lyman-$\alpha$ reconstruction informed by multiple high S/N observations. We scale the MUSCLES $\epsilon$ Eridani reconstruction by the ratio between the Lyman-$\alpha$ flux predicted by the \ion{Mg}{2} relation, $1.2 \times 10^{-13}$ erg cm$^{-2}$ s$^{-1}$, and the integrated Lyman-$\alpha$ flux reported by \citet{Youngblood:2016} for the $\epsilon$ Eridani reconstruction, $6.1 \times 10^{-11}$ erg cm$^{-2}$ s$^{-1}$. We replace the portion of the observed COS spectrum with a scaled version of the $\epsilon$ Eridani reconstruction in the interval 1214.63 -- 1216.78 \AA, where the boundaries are identified by the intersection points between the original observed spectrum and the scaled reconstruction. We assign errorbars that assume an uncertainty of a factor of 2 in either direction to be conservative. We expect that the true profile of V1298 Tau would have stronger pressure broadened wings, but most exoplanet applications of the Lyman-$\alpha$ flux for photochemistry are insensitive to the profile. If a reliable Lyman-$\alpha$ reconstruction for a closer analog to V1298 Tau becomes available in the future, we can update the data product accordingly. We also scale the $\epsilon$ Eridani MUSCLES spectrum to fill in the FUV detector gap of the SED from 1519.42 -- 1530.78 $\mathrm{AA}$, using the flux ratio of the nearby \ion{Si}{4} 1394/1403 \AA\ resonance doublet to determine the scaling factor in this spectral region.

\begin{deluxetable*}{cccc}
\tablecaption{Lyman-$\alpha$ Predictions From Correlations \label{tab:lya_corr}}
\tablehead{\colhead{Input Variable}&\colhead{Input Quantity}&\colhead{Predicted Lyman-$\alpha$}&\colhead{Reference}\\
\colhead{--}&\colhead{[various]}&\colhead{$ [10^{-13} \textrm{erg}\,\textrm{s}^{-1}\,\textrm{cm}^{-2}] $}&\colhead{--}}
\tabletypesize{\scriptsize}
\startdata
$\log_{10}$ $L_{\textrm{\ion{Si}{3} }}/L_{\textrm{bol}}$ & -5.61 & 1.0 & CA23\\
$\log_{10}$ $L_{\textrm{\ion{N}{5}}}/L_{\textrm{bol}}$ & -5.87 & 3.0 & CA23\\
Rossby Number & assumed saturation regime $ < \textrm{Ro}_c = 0.21$ & 6.8 & \citet{Pineda:2021a}\\
$\log_{10}$ \ion{Mg}{2} hk doublet Surface Flux  & 6.35 & 1.2 & \citet{Wood2005}\\
\enddata
\end{deluxetable*}

\section{Extreme Ultraviolet}
\label{sec:euv}
The EUV spectra of most stars are poorly constrained. The only facility to observe across this wavelength regime was the \emph{Extreme Ultraviolet Explorer} (\emph{EUVE}) which was operational from 1992 to 2001 and was not sensitive enough to obtain high signal-to-noise spectra for most main-sequence stars unless they were highly active and nearby. This has proven to be a significant obstacle to studying stellar magnetic activity and exoplanet atmospheric escape. In the absence of data for most stars, one must either rely on other observed quantities like the X-ray or Lyman-$\alpha$ flux and then use correlations between that quantity and the EUV flux of the few stars observed by EUVE \citep{Linsky:2014, Youngblood:2016, France:2020}, or use a model of the star's atmospheric structure above the photosphere \citep{Fontenla2016, Tilipman2020, Peacock2020}.

\subsection{Differential Emission Measure}
\label{sec:dem}
We use the differential emission measure (DEM) technique, described in detail in \citet{Duvvuri:2021} and variations of which have been used in a number of cases to estimate the XUV irradiation of exoplanets \citep{Sanz-Forcada:2004, Sanz-Forcada:2011, Louden:2017, Diamond-Lowe:2021, Diamond-Lowe:2022}, to estimate the extreme ultraviolet spectrum of V1298 Tau and fill in the gaps between observations. The DEM method uses observed emission to constrain the density and temperature structure of the upper stellar atmosphere expressed as a one-dimensional function of temperature $\Psi (T) = n_e n_{\rm{H}} \frac{ds}{dT}$ (i.e. the differential emission measure), and then combines this function with atomic data to predict unobserved emission produced from the same plasma that emitted the observed flux. The DEM function can be conceptually described as a collision or reaction rate for exciting electrons to higher states weighted by the amount of plasma along the line-of-sight at a given temperature \citep{Kashyap:1998, Craig:1976, Duvvuri:2021}. The intensity of a specific emission feature can be determined by using atomic data to construct its ``contribution function" (the energy contributed by this feature from an optically thin plasma at a particular temperature), weighting this function by the DEM, and then integrating over temperature. The peak of the integrand is the ``formation temperature" $T_{\textrm{formation}}$. To constrain the DEM, it is ideal to have measurements of multiple emission features that each have very narrowly peaked contribution functions to minimize the degeneracy of DEM shapes that could produce the observed emission, and whose formation temperatures densely occupy the full temperature range of interest ($10^4$ -- $10^8$ K for the stellar upper atmosphere).

We update the method described in \citet{Duvvuri:2021} by using a more recent version of \texttt{CHIANTI} (\texttt{v10.0.1, \citealt{Dere:1997, DelZanna:2021}}) and incorporating the recombination continua of hydrogen and helium species (this updated method was also used in \citealt{Feinstein:2022}). As described in \citet{Duvvuri:2021}, we use a 5th order Chebyshev polynomial to describe the functional form of $\log_{10} \Psi (T)$, assume the method has a parameterized intrinsic uncertainty that is a temperature-independent fraction $s$ of the predicted flux, and evaluate the likelihood of a given DEM function by directly comparing the observed line flux to the flux predicted by integrating the product of the DEM and contribution function in a Markov Chain Monte-Carlo (MCMC) sampler. Our approach differs from the iterative Monte-Carlo method \citep{Sanz-Forcada:2004} by allowing a greater range of ``acceptable" solutions; not just finding the ``best" DEM for a given Monte-Carlo sample of line flux distributions, but any DEM that produces a likely fit to the data. Our approach also differs from the more closely related method employed by \citet{Diamond-Lowe:2021} that used Chebyshev polynomials and MCMC sampling like \citet{Duvvuri:2021} but evaluated the likelihood in DEM-space, using the integral of the contribution function to determine an ``average DEM" value associated with each observed emission line and fitting to these averages, a method which has significant computational advantages but again restricts the range of allowed DEM shapes by neglecting the width and shape of the contribution function. We use the \texttt{emcee} \citep{Foreman-Mackey:2013} affine-invariant implementation of the Metropolis-Hastings MCMC algorithm \citep{goodman10} to sample the joint posterior distribution of the six Chebyshev polynomial coefficients and $s$-factor systematic uncertainty. We ran 25 chains for $2.2 \times 10^4 \lesssim 110 \tau$ steps, where $100 < \tau <  200$ steps is the range of autocorrelation times for all parameters calculated by \texttt{emcee}, and discard the first $2 \times 10^3$ steps from all walkers.

\begin{figure}
    \centering
    \includegraphics[width=\textwidth]{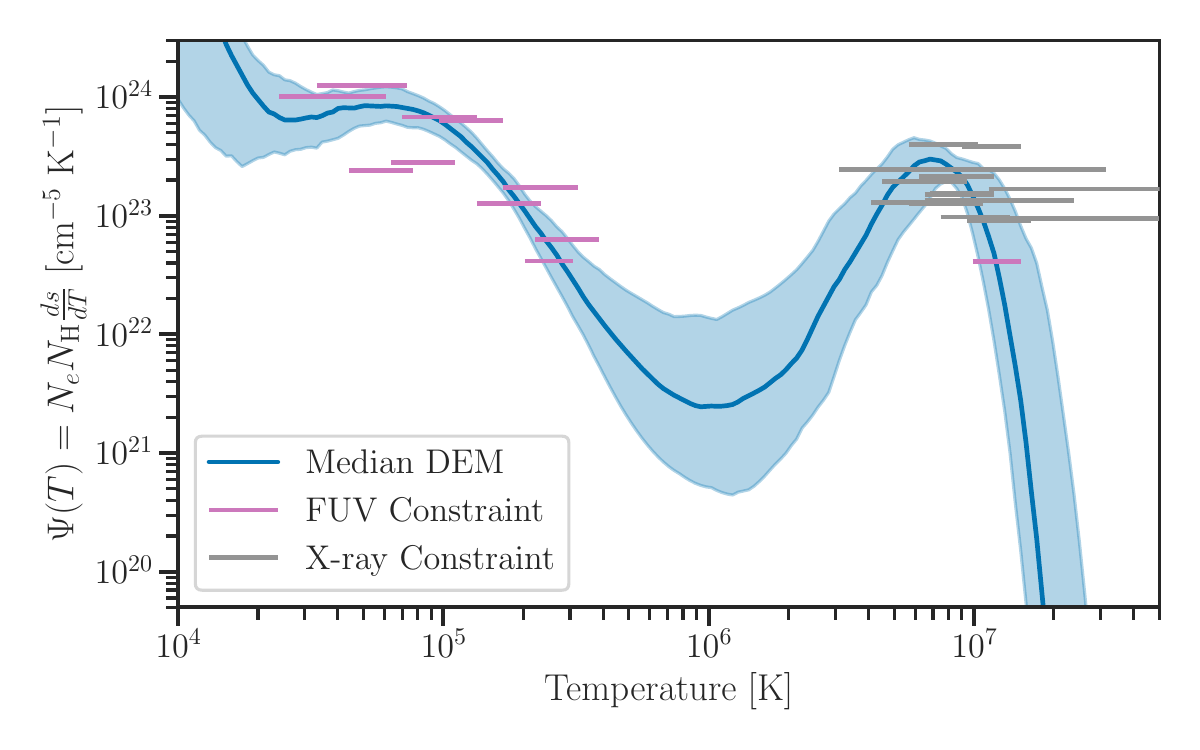}
    \caption{The Differential Emission Measure model fit compared to representative average DEM values derived from the observed fluxes used to constrain the fit. The uncertainty of allowed DEM shapes is greatest in the interval between $3 \times 10^5$ K -- $3 \times 10^6$ K where there are no observed emission features formed at specifically those temperatures. The peak at $6 \times 10^6$ K corresponds to the corona and the DEM turning down prevents the formation of emission lines at temperatures greater than $1.5 \times 10^7$ K, which is consistent with the isothermal XSPEC model fit to the X-ray data. Figure \ref{fig:compare} compares the fluxes predicted by the DEM model to the observed flux constraints.}
    \label{fig:dem}
\end{figure}

The X-ray spectral bins used to constrain the high-temperature end of the corona were selected by downsampling the spectral resolution of the XSPEC model spectrum to $R = \frac{\lambda}{\Delta \lambda} = 40$ to ensure all emission line profiles were contained within spectral bins, then identifying which bins had the highest integrals of their contribution functions. The chosen bins correspond to the strong emission lines between 0.7 and 1.1 keV shown in Figure \ref{fig:xray}, but each bin contains blends from multiple emission lines which cannot be resolved. The FUV constraints are more straightforward, the summed flux from observed emission lines of different species, with integrated fluxes from the line profile fits described in Section \S\ref{sec:analysis_fuv}, where we use lines that have not been significantly impacted by interstellar reddening. V1298 Tau is active enough that we were able to observe the \ion{Fe}{21} 1354 \AA\ coronal emission line, which provides a constraint at temperatures similar to the X-ray spectral bins and these appear to agree with each other.

\begin{figure}
    \centering
    \includegraphics[width=\textwidth]{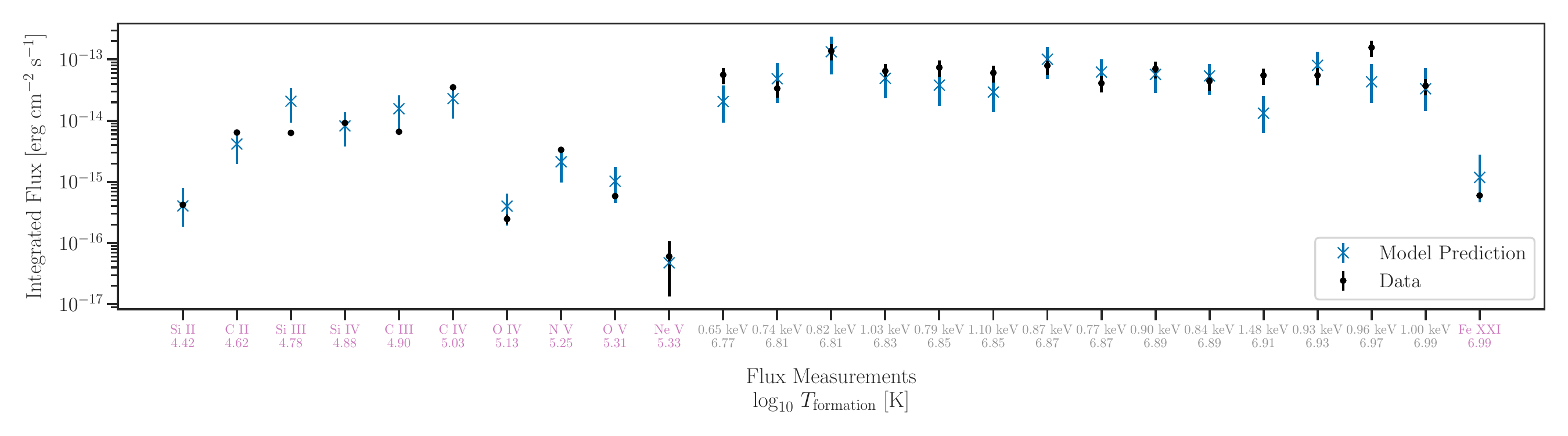}
    \caption{The observed flux constraints are plotted as black points with errorbars corresponding to their measurement uncertainties while the DEM model predictions are plotted as light blue crosses with errorbars corresponding to the 16$^{\textrm{th}}$ -- 84$^{\textrm{th}}$ percentile values of the distribution of fluxes predicted by drawing from the posterior of DEM shapes and the fractional flux systematic uncertainty parameter. The flux constraints are divided into two categories: ion species corresponding to integrated FUV emission line fluxes (labeled in pink) and central energies corresponding to the integrated flux of X-ray spectral bins (labeled in gray). Beneath each flux constraint's label is its $\log_{10} \left(T_{\textrm{formation}} \, [\textrm{K}]\right)$ value and the constraints are ordered by formation temperature increasing to the right.}
    \label{fig:compare}
\end{figure}

Figure \ref{fig:dem} shows the distribution of DEM shapes that fit the data, with the median DEM value represented by a solid blue line and the shaded region filling in the interval between the 16th and 84th percentile boundaries of DEM values returned by the sampled polynomial shapes. The horizontal lines represent constraints imposed by the observed fluxes, with the width encompassing the central 68\% of the cumulative integral of the contribution function and the $y-$value representing the average $\overline{\Psi}$ value obtained by dividing the flux by the integral of the contribution function (treating the DEM $\Psi$ as locally constant). These averages are illustrative and meant to show which temperatures are constrained by which measurements, color-coded to distinguish between the FUV lines (light pink) and X-ray spectral bins (gray). Figure \ref{fig:compare} compares the predicted fluxes from the DEM to the observed values and is a more direct visual representation of the model's goodness-of-fit while Table \ref{tab:lines} compares the observations and model predictions for all flux constraints used in the DEM-fitting process. As the width of the uncertainty swath in Figure \ref{fig:dem} indicates, the lack of observational constraints leads to high uncertainties at temperatures around $10^6$ K, the regime where the majority of EUV flux is formed. Direct observations of stellar EUV emission are necessary to reduce this uncertainty for any modeling approach.

The FUV and X-ray data were not taken simultaneously and if there were unresolved flares in either dataset the non-simultaneity would introduce discrepancies between the predicted EUV emission and the true quiescent spectrum of V1298 Tau. However, the good agreement between both X-ray observations indicates that they were at similar levels of flare activity, while no significant flares were noted in the FUV photon event lightcurve. The DEM average for the FUV Fe XXI line also agrees well with the constraints from the X-ray data, suggesting that any activity level discrepancies between these observations fall within the uncertainty of the measurements and fitting process.

\begin{deluxetable*}{ccccc}
\tablecaption{Integrated fluxes of optically thin FUV emission lines and X-ray spectral bins compared to the DEM predictions. \label{tab:lines}}
\tablehead{
\colhead{Emission Feature} & \colhead{Wavelengths} & \colhead{$\log_{10} T_\textrm{formation}$} & \colhead{Observed Flux} & {DEM Prediction}\\
\colhead{}    & \colhead{[$\textrm{\AA}$]}         & \colhead{$\log_{10}(\textrm{[K]})$}               & \colhead{[$10^{-15}$ erg\,s$^{-1}$\,cm$^{-2}$]} & \colhead{[$10^{-15}$ erg\,s$^{-1}$\,cm$^{-2}$]}}
\startdata
\ion{Si}{2} &  1260.4, 1264.7 &  4.42 &  $0.51 \pm 0.06$ & $ 0.40_{-0.22}^{+0.38}$\\
\ion{C}{2} &  1335 multiplet &  4.62 & $ 6.42 \pm  0.436$& $ 4.2_{-2.2}^{+2.7}$\\
\ion{Si}{3} &  1294.5, 1301.1&  4.78 & $6.53 \pm  0.314$ & $ 21_{-11}^{+13}$\\
\ion{Si}{4} & 1393.7, 1402.7 &  4.88 & $9.18 \pm  0.751 $&$  8.1_{-4.2}^{+5.4}$\\
\ion{C}{3} &  1175 multiplet &  4.90 & $6.6 \pm  0.314$ &$ 16_{-8.5}^{+10}$ \\
\ion{C}{4} &  1548.1, 1550.7&  5.03 & $35.1 \pm  3.02$ & $ 23_{-12}^{+14}$\\
\ion{O}{4} & 1401.1 &  5.13 & $0.247 \pm  0.044$ & $ 0.41_{-0.21}^{+0.24}$\\
\ion{N}{5} &  1238.8, 1242.8 &  5.25 & $3.34 \pm  0.244$ & $ 2.1_{-1.1}^{+1.3}$\\
\ion{O}{5} & 1371.3 &  5.31 & $0.587 \pm  0.05.84$ &$ 1.0_{-0.56}^{+0.76 }$\\
\ion{Ne}{5} & 1145.6 &  5.33 & $0.0604 \pm  0.047 $& $0.047_{-0.026}^{+0.040}$ \\
0.65 keV &  $ 19.1  \pm  0.31 $ &  6.77 & $56 \pm  17$ & $ 21_{-11}^{+16}$\\
0.74 keV &  $ 16.7  \pm  0.27 $ &  6.81 & $34 \pm  10$ & $ 49_{-28}^{+40}$\\
0.82 keV &  $ 15.2  \pm  0.25 $ &  6.81 & $140 \pm  41$ & $ 130_{-75}^{+100}$\\
1.03 keV &  $ 12.1  \pm  0.20 $ &  6.83 & $65 \pm  19$ & $ 50_{-26}^{+30}$\\
0.79 keV &  $ 15.7  \pm  0.26 $ &  6.85 & $74 \pm  22$ & $ 38_{-20}^{+25}$\\
1.10 keV &  $ 11.3  \pm  0.19 $ &  6.85 & $61 \pm  18$ & $ 29_{-15}^{+19}$\\
0.87 keV &  $ 14.2  \pm  0.21 $ &  6.87 & $79 \pm  24$ & $ 100_{-53}^{+60}$\\
0.77 keV &  $ 16.2  \pm  0.27 $ &  6.87 & $41 \pm  12$ & $63_{-33}^{+38}$\\
0.90 keV &  $ 13.7  \pm  0.15 $ &  6.89 &  $70 \pm  21$ &$ 58_{-30}^{+34}$ \\
0.84 keV &  $ 14.7  \pm  0.24 $ &  6.89 & $45 \pm  13 $& $54_{-27}^{+32}$ \\
1.48 keV &  $ 8.40  \pm  0.14 $ &  6.91 & $55 \pm  17$ & $13_{-7.0}^{+11}$\\
0.93 keV &  $ 13.3  \pm  0.22 $ &  6.93 & $55 \pm  17$ & $ 81_{-42}^{+54}$\\
0.96 keV &  $ 12.9  \pm  0.21 $ &  6.97 & $160 \pm  47$ & $ 43_{-24}^{+43}$\\
1.00 keV &  $ 12.5  \pm  0.20 $ &  6.99 & $37 \pm  11$ & $ 33_{-19}^{+38}$\\
\ion{Fe}{21} & 1354.0 &  6.99 & $0.598 \pm  0.0576$ & $1.2_{-0.70}^{+1.67}$ \\
\enddata
\tablecomments{In cases where multiple transitions are listed for the same ion, the reported flux is the summed flux across all listed transitions. For X-ray spectral bins, we list the central energy, wavelength, and wavelength bin width.}
\end{deluxetable*}

\subsection{EUV Spectrum}

As mentioned above, we have improved the method of \citet{Duvvuri:2021} to include recombination continua from hydrogen and helium species which adds bound-free edges, most notably the \ion{H}{1} recombination edge short of 912 \AA. In addition to propagating uncertainties with more specificity to all the observations of an individual star, an advantage of the DEM over scaling relations is the ability to synthesize an actual spectrum with higher wavelength resolution than the integrated flux across 100 \AA\ bandpasses. While the DEM cannot predict line profiles, predicting the flux from individual optically thin emission lines allows spectral synthesis at a resolution where the width of a line is contained within a single spectral bin. This is especially important for modelling atmospheric escape from the exospheres of irradiated exoplanets with methods more sophisticated than energy-limited escape. As observations of the \ion{He}{1} 10830 \AA\ line become increasingly accessible for exoplanets, \citet{Oklopcic:2019} demonstrates the necessity of well-characterized EUV and mid-UV spectra with uncertainties to interpret those observations.

\begin{figure}
    \centering
    \includegraphics[width=\textwidth]{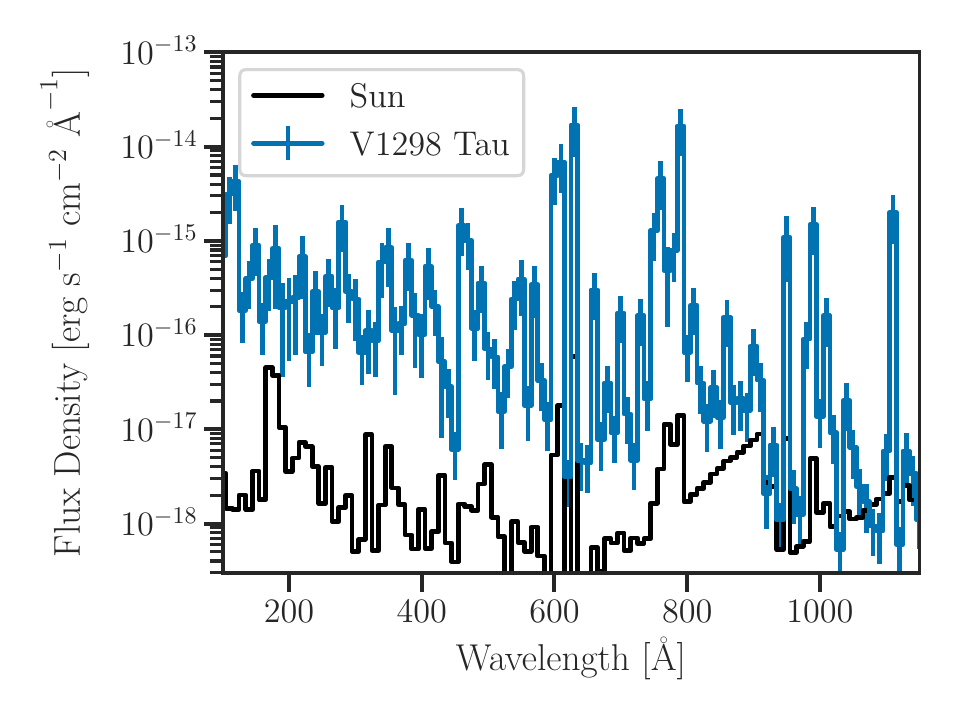}
    \caption{The EUV spectrum of V1298 Tau (light blue) compared to the EUV spectrum of the quiescent Sun \citep{Woods:2009}. The EUV spectrum of younger, more active V1298 Tau is consistently a factor of 100 -- 1000 greater than the Sun's across this wavelength regime, with a shallower slope for the \ion{H}{1} continuum blueward of 912 \AA forming the base of the strong emission lines.\label{fig:euv}}
\end{figure}

One set of parameters from the posterior distribution describes the shape of the DEM and the intrinsic uncertainty on fluxes predicted by that DEM. For each sample draw from the posterior we calculate $\Psi$ using the Chebyshev coefficients, predict the flux $f$ in 1 \r{A} bins from 1 to 2000 \r{A} using the contribution functions of all lines that \texttt{CHIANTI} lists within the wavelength bin, and then sample from a Gaussian $\mathcal{N}(\mu = f, \sigma = s \cdot f)$ where $s$ is the fractional systematic uncertainty parameter. This creates one spectrum output corresponding to the single draw of parameters from the posterior distribution. After $10^6$ such draws we record the 16th, 50th, and 84th percentile values of the flux in each wavelength bin to infer the EUV spectrum and the uncertainty of the inference. Figure \ref{fig:euv} shows the EUV portion of the predicted spectrum compared to the Solar Irradiance Reference Spectrum from \citet{Woods:2009} scaled to the distance from V1298 Tau, illustrating how youth and activity enhance the flux of V1298 Tau across the entire EUV regime. The integrated XUV (X-ray + EUV, $< 912$ \AA) flux from V1298 Tau using our combination of the XSPEC model and the DEM-generated EUV spectra is $3.2 \pm 0.3 \times 10^{-12}$ erg s$^{-1}$ cm$^{-2}$ with additional uncertainty scaling factors of 15\% and 20\% introduced to the SED for the FUV flux calibration and $n($\ion{H}{1}$)$ column density uncertainties respectively.

\citet{Poppenhaeger:2021} and more recently \citet{Maggio:2023} have also estimated the total EUV flux of V1298 Tau using different sets of observations and methods than this work. Between all three sets of observations, it is clear in the data that V1298 Tau exhibits significant long-term X-ray variability, but assessing the EUV variability is more difficult given the model-dependence of the EUV estimation. The X-ray fluxes reported by \citet{Poppenhaeger:2021}, \citet{Maggio:2023}, and this work are listed in Table \ref{tab:xray_comparison}.

\begin{deluxetable}{cccc}
\tablecaption{X-ray fluxes reported across three sets of observations from \citet{Poppenhaeger:2021}, \citet{Maggio:2023}, and this work. The best-fit models indicate that the coronal flux of V1298 Tau has varied by a factor of 2 between 2019 -- 2021 while the intra-observation variability has been $< 30\%$. \label{tab:xray_comparison}}

\tablehead{
\colhead{Reference} & \colhead{X-ray Telescope} & \colhead{Observation Period} & \colhead{Best-fit Model Unabsorbed $F_{0.1 \text{--} 2.4 \, \mathrm{keV}}$}\\
\colhead{} & \colhead{} & \colhead{} & \colhead{[$10^{-12}$ erg\,s$^{-1}$\,cm$^{-2}$]}}

\startdata
\citet{Poppenhaeger:2021} & \emph{ROSAT} + \emph{Chandra} & 1991 + November 2019 & $0.92 \pm 0.1$ \\
This work & \emph{NICER} & October/November 2020 & $1.74 \pm 0.025$ \\
\citet{Maggio:2023} & \emph{XMM}-\emph{Newton} & August 2021 (quiescent) & $1.4^{+ 0.1}_{-0.2}$ \\
\citet{Maggio:2023} & \emph{XMM}-\emph{Newton} & August 2021 (elevated) & $1.82^{+ 0.03}_{-0.08}$
\enddata

\end{deluxetable}

\citet{Maggio:2023} fits emission measure distributions to two sets of observations with different X-ray fluxes labelled ``quiescent" and ``elevated", finding that the bulk of the difference in X-ray flux can be attributed to the enhancement of a hotter 10$^7$ K plasma component in the elevated state. This would likely have a small impact on the EUV variability since the majority of EUV flux is formed between $10^{5.5}$ and $10^{6.5}$ \citep{Duvvuri:2021}. If there is significant EUV variability in this system, either between or during observations, it will affect both the detection of atmospheric escape and the inference of mass-loss rates via transmission spectroscopy, and this possibility should be considered in future analyses of planets in this system.

\section{Conclusion}
\label{sec:conclusion}

As the star spins down, the non-thermal heating of the star's upper atmosphere will decrease over time and reduce the high-energy emission from V1298 Tau, but not necessarily by a constant value across the XUV wavelength regime depending on how the evolution varies at different stellar atmospheric heights and temperatures \citep{Ribas:2005}. The long-term fate of V1298 Tau's planets will depend on how the photoevaporative mass-loss changes over the lifetime of the system. \citet{Ribas:2005} assembled spectra of 7 solar-mass stars (0.9 -- 1.1 $M_{\odot}$) across a wide range of ages, including \emph{EUVE} data, to characterize these stars' evolution of high-energy emission over time. \citet{Ribas:2005} fit power-laws to the integrated flux for 3 XUV bandpasses: 1 -- 20, 20 -- 100, and 100 -- 360 \AA\ and assigned a power-law for the 360 -- 920 \AA\ bandpass. More recent work like \cite{Wright:2011} has favored a broken power-law for X-ray emission, observing that for the youngest stars, the X-ray emission clusters around a saturation value, well below what the \citet{Ribas:2005} power-laws would predict if allowed to extend to those young ages.

\begin{figure}
    \centering
    \includegraphics[width=\textwidth]{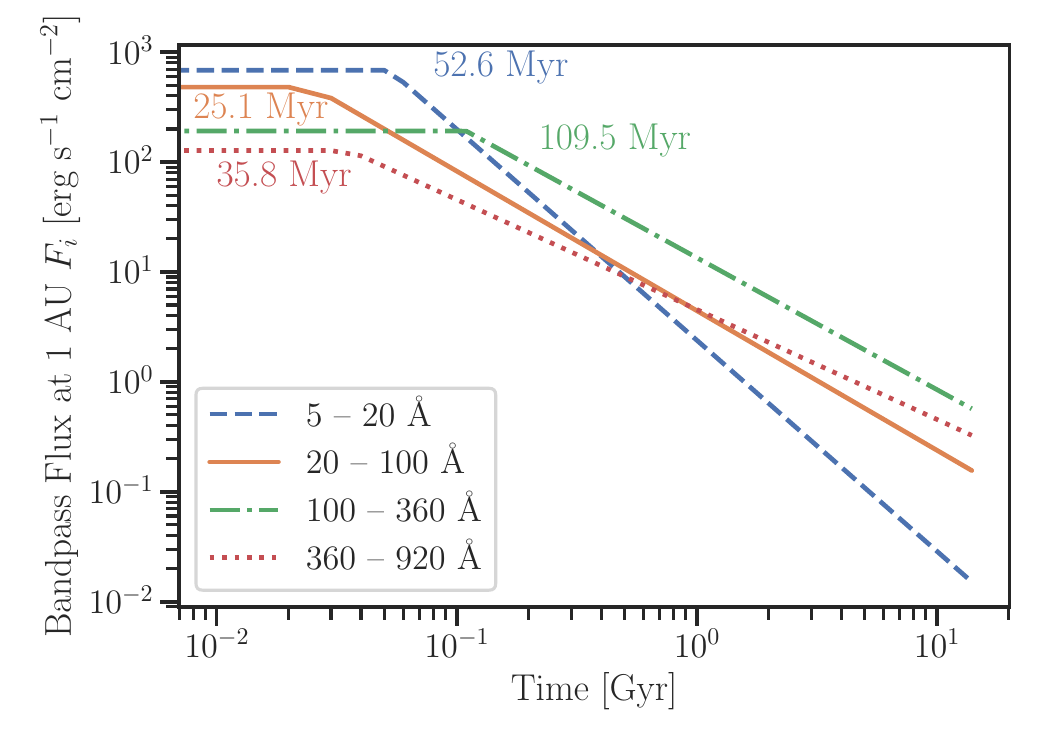}
    \caption{The broken-power laws describing the evolution of high-energy emission for solar-mass stars divided into 4 bandpasses, annotated with the time corresponding to the breakpoint of the power-law: 1 -- 20 \AA\ (dashed dark blue, 52.9 Myr), 20 -- 100 \AA\ (solid orange, 24.2 Myr), 100 -- 360 \AA\ (dot-dashed green, 108.7 Myr), 360 -- 920 \AA\ (dotted red, 35.9 Myr). The parameters for the broken power-laws are listed in Table \ref{tab:power-law}.}
    \label{fig:power-law}
\end{figure}

V1298 Tau is not a perfect Young Sun analog, but the original planet discovery paper, \citet{David:2019a}, estimated that V1298 Tau would settle close to either side of the F/G cusp. Tables 5 and 6 from \citet{Pecaut:2013}\footnote{updated version hosted at: \url{https://www.pas.rochester.edu/~emamajek/EEM_dwarf_UBVIJHK_colors_Teff.txt}} predict that a star of this mass will settle on the main-sequence as a $T_{\mathrm{eff}} = 6000$ K F9 -- F9.5V star, and V1298 Tau is old enough that its mass should not change significantly during that process. This would make the future main-sequence V1298 Tau very similar to $\beta$ Comae Berenices, the hottest star in the \citet{Ribas:2005} sample (G0V, $T_{\mathrm{eff}} = 6000$ K, $M_\star = 1.1 \, M_{\odot}$), used to anchor the power-law relations at 1.6 Gyr. By taking V1298 Tau to be representative of the saturation flux for young solar-mass stars, we modify the \citet{Ribas:2005} power-laws to be broken power-laws that follow

\begin{equation}
    F_i = \left\{ \begin{array}{lr}
        F_{\textrm{V1298 Tau}, i}, & \text{if } t < t_{\textrm{crit}, i}\\
        \alpha_i \left(\frac{t}{\textrm{1 Gyr}}\right)^{\beta_i} & \text{if } t \geq t_{\textrm{crit}, i}
    \end{array}
    \right\}
\end{equation}
\noindent where $i$ represents the individual bandpass intervals, $F_{\textrm{V1298 Tau}, i}$ is the flux of V1298 Tau scaled to a distance of 1 AU and integrated over the bandpass $i$, $\alpha_i$ and $\beta_i$ are taken from Table 5 of \citet{Ribas:2005}, and we solve for the breakpoint of the power-law $t_{\textrm{crit}, i} = \sqrt[\beta_i]{\frac{F_{\textrm{V1298 Tau}, i}}{\alpha_i}}$ by requiring the function to be continuous. The parameters for this broken-power law are listed in Table \ref{tab:power-law} and the functions are plotted in Figure \ref{fig:power-law}. The reported uncertainties on $t_{\textrm{crit}}$ only incorporate the uncertainty of the V1298 Tau SED and are therefore underestimates: the \citet{Ribas:2005} power-laws were calibrated with only one Solar analog at each representative age of their sample, and the 360 -- 920 $\textrm{\AA}$ bandpass was only anchored by the Sun and an assumed power-law slope. Determining the true evolution of this EUV bandpass is important for characterizing atmospheric escape and the relationship between spin-down and weakening stellar magnetism. \citet{Ribas:2005} notes that the power-law slopes grow shallower with increasing bandpass wavelengths, a trend in agreement with the finding from \citet{Ayres:1999} that the emission from hotter plasma decays more rapidly. A related observation from \citet{Pineda:2021a} is that the $t_{\textrm{crit}}$ values for broken power-laws fit to rotation-age-activity relations from FUV emission lines (transition region) are later than those derived from X-ray emission (corona). This work's broken power-law for the 360 -- 920 \AA\ bandpass diverges significantly from these findings in the literature, but is also the least constrained by data. Observations of the multiwavelength behavior of both the decay slope and breakpoint from activity saturation would be powerful tests for physical models of stellar magnetic evolution.

\begin{deluxetable*}{ccccc}
\tablecaption{Broken power-laws describing the evolution of bandpass fluxes for solar-type stars determined by linking the \citet{Ribas:2005} relations to the V1298 Tau SED collated in this work.\label{tab:power-law}}
\tablehead{
\colhead{Bandpass $i$} & \colhead{Flux at 1 AU $F_{\textrm{V1298 Tau}, i}$} & \colhead{$^\star \alpha_i$} & \colhead{$^\star \beta_i$} & {$t_{\textrm{crit}, i}$}\\
\colhead{[$\textrm{\AA}$]}    & \colhead{[erg\,s$^{-1}$\,cm$^{-2}$]} & \colhead{[erg\,s$^{-1}$\,cm$^{-2}$]}               & \colhead{[--]} & \colhead{[Myr]}}
\startdata
1 -- 20 & 685 & 2.4 & -1.92 & 53 $\pm 1$ \\
20 -- 100 & 480 & 4.45 & -1.27 & 25 $\pm 1$\\
100 -- 360 & 192 & 13.5 & -1.2 & 110 $\pm 30$\\
$^{\dagger}$360 -- 920 & 127 & 4.56 & -1 & 36 $\pm 7$ \\
\enddata
\tablenotetext{\star}{Table 5 of \citet{Ribas:2005}}
\tablenotetext{^\dagger}{No data for stars other than the Sun were available for this bandpass so \cite{Ribas:2005} calibrated the power-law by assuming $\beta = -1$ and solving for $\alpha$ to match the observed flux from the Sun.}
\end{deluxetable*}

\begin{figure}
    \centering
    \includegraphics[width=\textwidth]{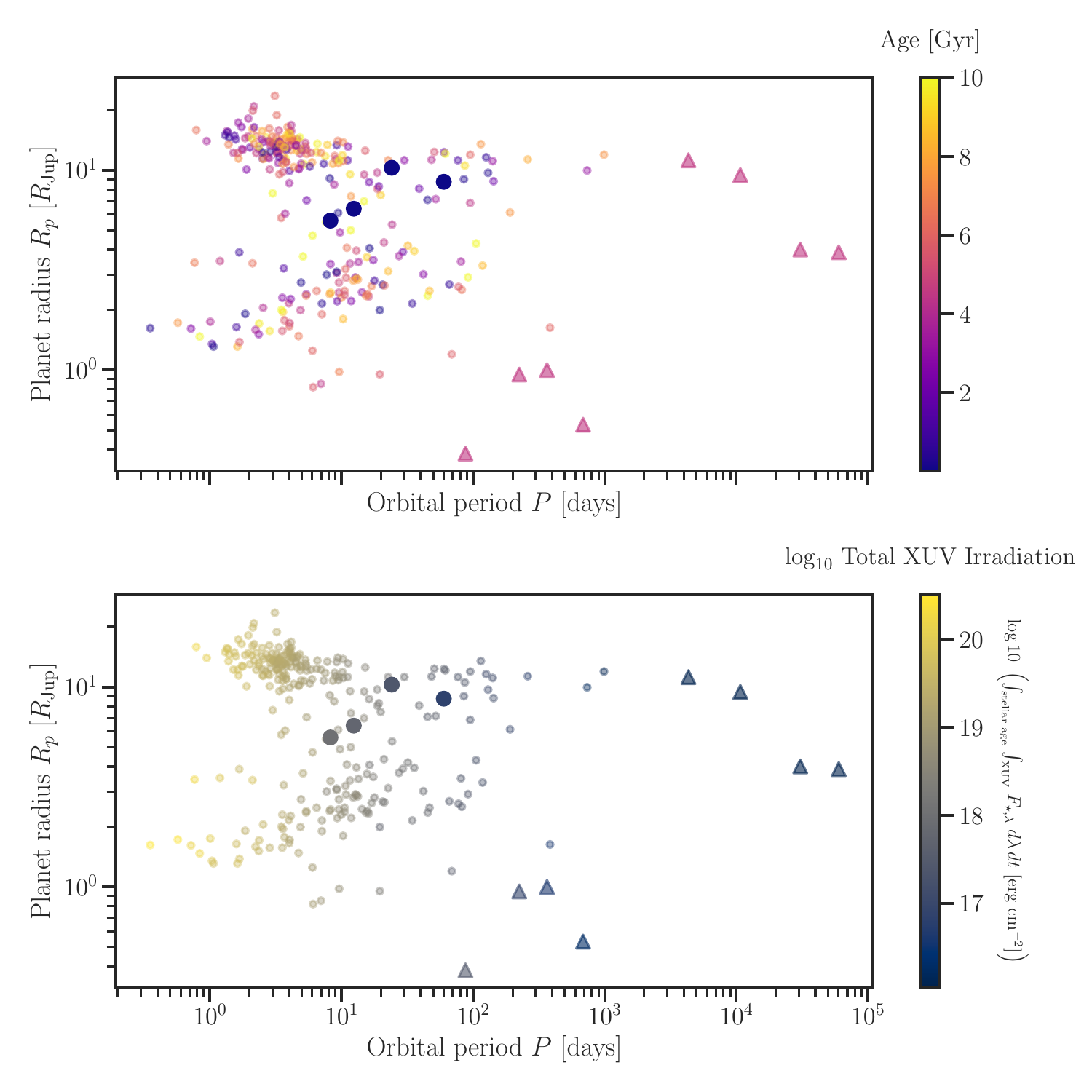}
    \caption{Both panels plot the planet radius against the orbital period for: a sample of confirmed exoplanets orbiting stars with a mass similar to V1298 Tau (translucent dots), the V1298 Tau planets (opaque circles), and Solar System planets (opaque triangles). The top panel colors the planet markers by the age of the star, with darker shades representing young systems and increasing brightness with age, while the bottom panel colors the planet markers by the cumulative XUV irradiation experienced by the planet ($F_\star$ is the flux received by the planet) assuming it has stayed at its current orbit for the entirety of the system's age, with the brightness of the color increasing with irradiation. For this sample selected by stellar mass, where all plotted planets are assumed to have experienced the same high-energy evolution, the cumulative XUV irradiation is a function of age. In a broader sample, where different stellar hosts follow different XUV irradiation evolution behavior, the cumulative XUV irradiation will also depend on other parameters like stellar mass.}
    \label{fig:xuv_population}
\end{figure}

The combination of transit surveys and \emph{Gaia} has made it possible to identify exoplanet systems in moving groups and associations with known ages, increasing the number of systems with precisely known ages. We queried the Exoplanet Archive\footnote{\url{https://exoplanetarchive.ipac.caltech.edu/}} for all confirmed exoplanets with known radii and orbital periods orbiting stars with $0.9 < M_\star < 1.2 M_\odot$ (similar to V1298 Tau $M_\star = 1.1 M_\odot$) and a reported age with an uncertainty less than a factor of 2, then applied the broken power-law evolution to each planetary system to determine the cumulative XUV irradiation of each planet (flux received by the planet integrated over XUV wavelengths and the lifetime of the system), plotted in Figure \ref{fig:xuv_population}. The planets of the V1298 Tau system are in a relatively sparse region of the plot, but there are a wide range of ages and XUV irradiation values represented amongst these planets' closest neighbors, with fairly little variation of total irradiation across planets near a particular orbital period. This is because of the rapid decay of XUV emission past 0.1 Gyr for solar-mass stars, leading to little difference in the cumulative irradiation for all but the youngest exoplanets orbiting this spectral type. However, the relatively later and slower decay of XUV emission from cooler exoplanet hosts \citep{Linsky2020} will complicate the dominance of orbital period in a more mixed sample of exoplanets. Looking for trends in XUV irradiation and planet demographics will require filling out this plot and others like it with different planetary parameters by increasing the range of stellar types with well-characterized XUV evolution.

As exoplanet surveys continue to detect viable systems for atmospheric characterization via transmission spectroscopy and direct-imaging, interpreting these observations and studying atmospheric evolution requires more detailed stellar characterization beyond spectral type. V1298 Tau is one of the brightest exoplanet hosts accessible within our solar neighborhood ($d = 108.5 pc$) and we still require model-dependent estimates of its high-energy emission. This star is an unusual case where the EUV uncertainties are more tightly constrained than the Lyman-$\alpha$ recovery, but both wavelength regimes need next-generation observatories to improve our understanding of stellar magnetism and the evolution of exoplanet atmospheres. This paper presents a roadmap for calculating empirically-informed spectra of exoplanet host stars that can be used until those observatories become available.

\bigskip\bigskip \bigskip\bigskip
\section*{Acknowledgements}
We thank their anonymous referee for their comments which clarified and improved the discussion of this work. This research has made use of data and software provided by the High Energy Astrophysics Science Archive Research Center (HEASARC), which is a service of the Astrophysics Science Division at NASA/GSFC, made use of  software provided by the Chandra X-ray Center (CXC) in the application packages CIAO and Sherpa, and made use of the NASA Exoplanet Archive, which is operated by the California Institute of Technology, under contract with the National Aeronautics and Space Administration under the Exoplanet Exploration Program. This work used atomic data from CHIANTI, a collaborative project involving George Mason University, the University of Michigan (USA), University of Cambridge (UK) and NASA Goddard Space Flight Center (USA). This work is partially based upon effort supported by NASA under award 80NSSC22K0076 and was supported by grant HST-GO-16163 to the University of Colorado. All the {\it HST} data used in this paper can be found in MAST: \dataset[10.17909/v5c1-4563]{http://dx.doi.org/10.17909/v5c1-4563}.


\facilities{
HST: STIS \citep{Woodgate:1998},
HST: COS \citep{Green:2012}),
NICER \citep{Gendreau:2016}}

\software{
Astropy \citep{AstropyCollaboration:2013, AstropyCollaboration:2018, AstropyCollaboration:2022},
bibmanager \citep{bibmanager:2020},
CHIANTI \citep{Dere:1997, DelZanna:2021},
CIAO \citep{CIAO:2001},
HEASoft \citep{HEASARC:2014},
emcee \citep{Foreman-Mackey:2013},
matplotlib \citep{matplotlib:2007},
numpy \citep{numpy:2020},
pandas \citep{pandas:2010, pandas:2020},
seaborn \citep{seaborn:2021},
XSPEC \citep{Arnaud:1996},
}

\bibliography{references}{}
\bibliographystyle{aasjournal}

\end{document}